\documentclass{nature}
\usepackage[utf8]{inputenc}
\usepackage{amsmath} 
\usepackage{amssymb} 
\usepackage{amsthm} 
\usepackage{array} 
\usepackage{bm} 
\usepackage{empheq}
\usepackage{float} 
\usepackage{mathtools} 
\usepackage[version=4]{mhchem} 
\usepackage{multirow}
\usepackage{siunitx} 
\usepackage{tensor}
\usepackage{textcomp} 
\usepackage[hyphens]{url}
\usepackage{xcolor}
\usepackage{xr-hyper}

\usepackage[final]{pdfpages} 

\usepackage{hyperref} 
\hypersetup{
    colorlinks=true,
    linkcolor=black,
    filecolor=cyan,      
    urlcolor=magenta,
    citecolor=black,
    pdftitle={Sharelatex Example},
    pdfpagemode=FullScreen
}

\usepackage{graphicx} 
\usepackage{subcaption} 

\usepackage{listings}
\definecolor{codegreen}{rgb}{0,0.6,0}
\definecolor{codegray}{rgb}{0.5,0.5,0.5}
\definecolor{codepurple}{rgb}{0.58,0,0.82}
\definecolor{backcolour}{rgb}{0.95,0.95,0.92}

\lstdefinestyle{mystyle}{
    backgroundcolor=\color{backcolour},   
    commentstyle=\color{codegreen},
    keywordstyle=\color{magenta},
    numberstyle=\tiny\color{codegray},
    stringstyle=\color{codepurple},
    basicstyle=\ttfamily\footnotesize,
    breakatwhitespace=false,         
    breaklines=true,                 
    captionpos=b,                    
    keepspaces=true,                 
    numbers=left,                    
    numbersep=5pt,                  
    showspaces=false,                
    showstringspaces=false,
    showtabs=false,                  
    tabsize=2
}
\usepackage{graphicx}
\usepackage{amsmath}
\usepackage{amssymb}
\usepackage{textcomp}
\usepackage{gensymb}
\usepackage{xcolor}
\usepackage{lineno}
\usepackage{soul}
\usepackage{colortbl}
\usepackage{ulem}
\usepackage[labelfont=bf]{caption}
\usepackage{multibib}
\newcites{supp}{Supplementary References}
\makeatletter
\let\saved@includegraphics\includegraphics
\AtBeginDocument{\let\includegraphics\saved@includegraphics}
\renewenvironment*{figure}{\@float{figure}}{\end@float}
\makeatother
\newcommand{\reftextit}[1]{}
\title{\large An active metasurface enhanced with moir\'e ferroelectricity}

\author{Dong Seob Kim$^{1,2}$, Chengxin Xiao$^{3}$, Roy C. Dominguez$^{4}$, Zhida Liu$^{1,2}$, Hamza Abudayyeh$^{1,2}$, Kyoungpyo Lee$^{1,2}$, Rigo Mayorga-Luna$^{4}$, Hyunsue Kim$^{1,2}$, Kenji Watanabe$^{5}$, Takashi Taniguchi$^{6}$, Chih-Kang Shih$^{1,2}$, Yoichi Miyahara$^{4,7}$, Wang Yao$^{3}$ and Xiaoqin Li$^{1,2*}$}

\begin{document}
\maketitle

{\renewcommand{\baselinestretch}{1.5}

\begin{affiliations}
    \item Department of Physics and Center for Complex Quantum Systems, The University of Texas at Austin, Austin, Texas, 78712, USA.
    \item Center for Dynamics and Control of Materials and Texas Materials Institute, The University of Texas at Austin, Austin, Texas, 78712, USA.
    \item New Cornerstone Science Laboratory, Department of Physics, The University of Hong Kong, Hong Kong, China.
    \item Department of Physics, Texas State University, San Marcos, Texas, 78666, USA.
    \item Research Center for Electronic and Optical Materials, National Institute for Materials Science, 1-1 Namiki, Tsukuba 305-0044, Japan.
    \item Research Center for Materials Nanoarchitectonics, National Institute for Materials Science, 1-1 Namiki, Tsukuba 305-0044, Japan.
    \item Materials Science, Engineering and Commercialization Program (MSEC), Texas State University, San Marcos, Texas, 78666, USA.

    \thanks{
  Corresponding authors:\textcolor{blue}{{elaineli@physics.utexas.edu}}}
\end{affiliations}
}

{\renewcommand{\baselinestretch}{1.7}

\date{\today}
\pagebreak

\begin{abstract}
Semiconductor moir\'e systems, characterized by their periodic spatial light emission, unveil a new paradigm of active metasurfaces. Here, we show that ferroelectric moir\'e domains formed in a twisted hexagonal boron nitride (t-hBN) substrate can modulate light emission from an adjacent semiconductor MoSe$_2$ monolayer, enhancing its functionality as an active metasurface. The electrostatic potential at the surface of the t-hBN substrate provides a simple way to confine excitons in the MoSe$_2$ monolayer. The excitons confined within the domains and at the domain walls are spectrally separated due to a pronounced Stark shift. Moreover, the patterned light emission can be dynamically controlled by electrically gating the ferroelectric domains, introducing a novel functionality beyond conventional semiconductor moir\'e systems. Our findings chart an exciting pathway for integrating nanometer-scale moir\'e ferroelectric domains with various optically active functional layers, paving the way for advanced nanophotonic applications.
\end{abstract}
}

\maketitle
\renewcommand{\baselinestretch}{2}
\setlength{\parskip}{7pt}

While semiconductor monolayers such as transition metal dichalcogenides (TMDs) have already found many photonic applications such as optical modulators~\cite{seyler_electrical_2015, sun_optical_2016}, detectors~\cite{lopez-sanchez_ultrasensitive_2013, baugher_optoelectronic_2014}, and light-emitting devices~\cite{ross_electrically_2014, withers_light-emitting_2015}, a moir\'e superlattice consisting of a twisted bilayer offers a significantly expanded parameter space for tunable properties. With sub-wavelength periodic light emission patterns, semiconductor moir\'e superlattices are emerging as an innovative form of active metasurfaces.   They feature prominent exciton resonances whose resonant energies, lifetimes, and transport all depend on the twist angle~\cite{wilson_excitons_2021, huang_excitons_2022, mak_semiconductor_2022}. However, many properties of moir\'e excitons change simultaneously, imposing undesirable constraints. For example, as the lateral confinement size reduces with an increasing twist angle, the exciton lifetime increases rapidly, likely leading to a reduced quantum efficiency~\cite{choi_moire_2020, choi_twist_2021}.

A different class of van der Waals materials, hexagonal boron nitride (hBN) is often used as passive encapsulation layers or tunnel barriers. It has become increasingly relevant as a photonic material for hosting quantum defect emitters~\cite{tran_quantum_2016, grosso_tunable_2017}, detecting UV light~\cite{maity_hexagonal_2018}, or as a natural hyperbolic material in the mid-infrared range~\cite{dai_tunable_2014, caldwell_sub-diffractional_2014}. Very recently, slid or twisted hBN (t-hBN) bilayers with a parallel interface have been found to exhibit ferroelectricity~\cite{yasuda_stacking-engineered_2021, vizner_stern_interfacial_2021}. Furthermore, it has been suggested that the periodic electrostatic potential on the top t-hBN surface can alter the properties of an adjacent functional layer placed on top of the t-hBN substrate~\cite{zhao_universal_2021}. In this case, the generation of the moir\'e potential is separated from the functional layer, offering greater flexibility.

Here, we explore a new strategy to create active metasurfaces by seamlessly integrating the ferroelectric (FE) functionality of a t-hBN with a light-emitting semiconductor monolayer. 
The electrostatic potential on the surface of the t-hBN confines excitons in an adjacent MoSe$_2$ monolayer. Those excitons confined within the domains and along the domain walls (DWs) are spectrally separated due to the Stark shift induced by an in-plane electric field strongest at the DWs. By correlating Kevin Probe Force Microscopy (KPFM) and optical microscopy experiments, we show that the spatial light emission pattern follows the moir\'e pattern of the t-hBN substrate. Through electric gating, one can erase and restore the FE domains. Consequently, the light emission from the MoSe$_2$ monolayer exhibits characteristic hysteresis behavior. Since the FE domain size is readily controlled by the twist angle, the light emission pattern can be modulated at a length scale far below the optical diffraction limit, charting a new pathway for creating active metasurfaces~\cite{wu_programmable_2020, forbes_structured_2021}.

\begin{figure}[H]
    \centering
    \includegraphics[width=16cm]{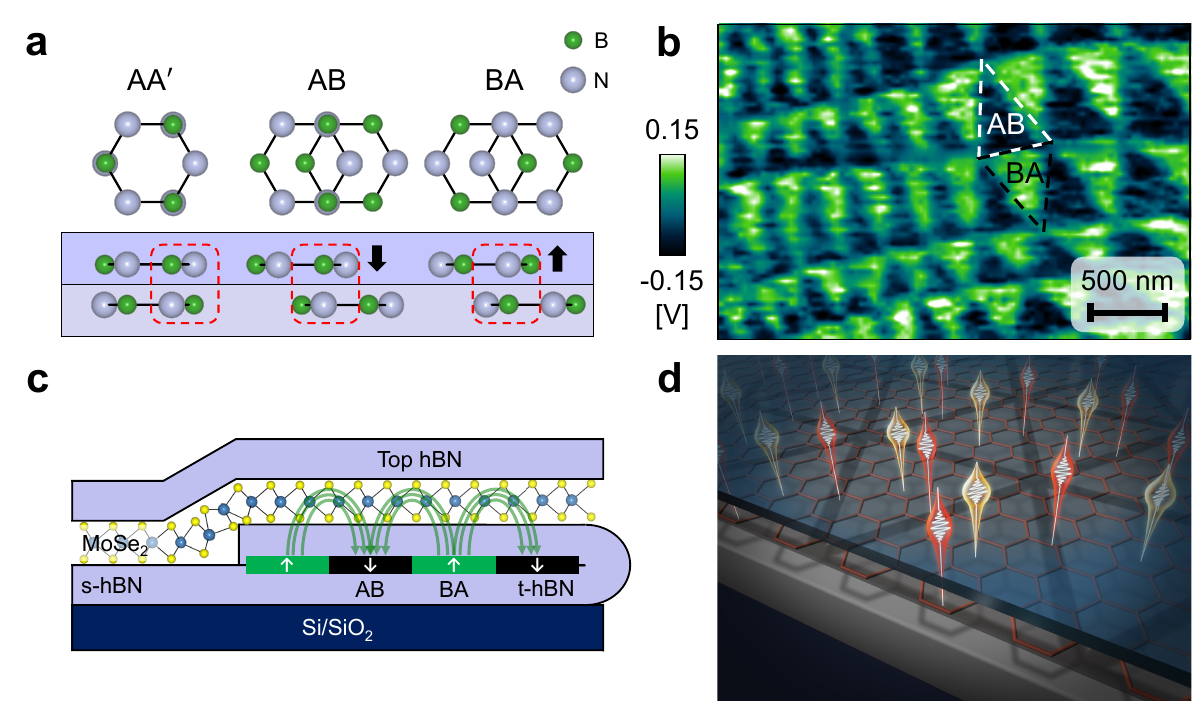}
    \caption{\textbf{Moir\'e ferroelectric domains in t-hBN modulate light emission from an adjacent semiconductor layer.}
    \textbf{a}, Top and side views of the hBN parallel interface at high-symmetry points. Inversion symmetry breaking leads to spontaneous polarization as indicated by down (up) black arrows corresponding to AB (BA) stacking.
    \textbf{b}, KPFM image of the electrostatic moir\'e potential on the top surface of a t-hBN substrate. AB (BA) domains are marked by white (black) dashed triangles.
    \textbf{c}, Sketch of a t-hBN/MoSe$_2$ monolayer/hBN vdW multilayer with alternating domains at the twisted hBN interface. Green lines and arrows illustrate the E-field generated by the FE domains with the largest in-plane E-field at the DWs.
    \textbf{d}, Illustration of spectral and spatial modulation of light emission from a semiconductor functional layer on top of a t-hBN substrate}
    \label{fig:fig1}
\end{figure}

We first explain the formation of FE domains in a t-hBN bilayer conceptually as depicted in Fig.~\ref{fig:fig1}a. A natural hBN crystal consists of layers stacked in AA' sequence, in which hexagonal lattices overlap, and B (N) atoms are vertically aligned with corresponding N (B) atoms in adjacent layers. In another energetically favorable AB or BA (Bernal) stacking configuration, in which hexagonal lattices laterally slide, the B (N) atoms in the upper layer align with the N (B) atoms in the lower layer, and the N (B) atoms in the upper layer are positioned above the vacant center of the hexagon in the lower layer. The spatial inversion symmetry is broken, leading to opposite polarization directions between AB and BA stacking as indicated by the black arrows.

An array of FE domains with alternating polarization directions are formed in a t-hBN bilayer in which the domain size is readily controlled by the twist angle. These FE domains can be directly visualized via KPFM shown in Fig.~\ref{fig:fig1}b as an example. Details of the sample preparation and KPFM measurements can be found in Methods. The color contrast in a KPFM image represents the electrostatic potential near the top surface of the t-hBN, which is related to polarization via
\begin{equation}
    V(\mathbf{R}, z)
    \approx \text{sgn}(z)\frac{P(\mathbf{R})}{2\epsilon_{0}}e^{-G|z|},
    \label{eqn:eqn1}
\end{equation}
where the net polarization is calculated from $P(\mathbf{R})=\int z'\Delta\rho(\mathbf{R},z')dz'$, $G=\frac{4\pi}{\sqrt{3}b}$, $\mathbf{R}$ represents the lateral position vector, $b$ characterizes the supercell size, and $z$ is the vertical distance to the buried interface~\cite{zhao_universal_2021}. The magnitude of $P$ is 2.01 pCm$^{-1}$ obtained from the first principles calculations. The surface potential extracted from KPFM measurements agrees remarkably well with the prediction of this theory~\cite{kim_electrostatic_2024}. When a semiconductor functional layer is placed on top of a t-hBN substrate, the E-field (green lines in Fig.~\ref{fig:fig1}c) generated by the surface potential of the t-hBN can periodically modulate the spectral and spatial pattern of the semiconductor layer as illustrated in Fig.~\ref{fig:fig1}d. 

\begin{figure}[H]
    \centering
    \includegraphics[width=\linewidth]{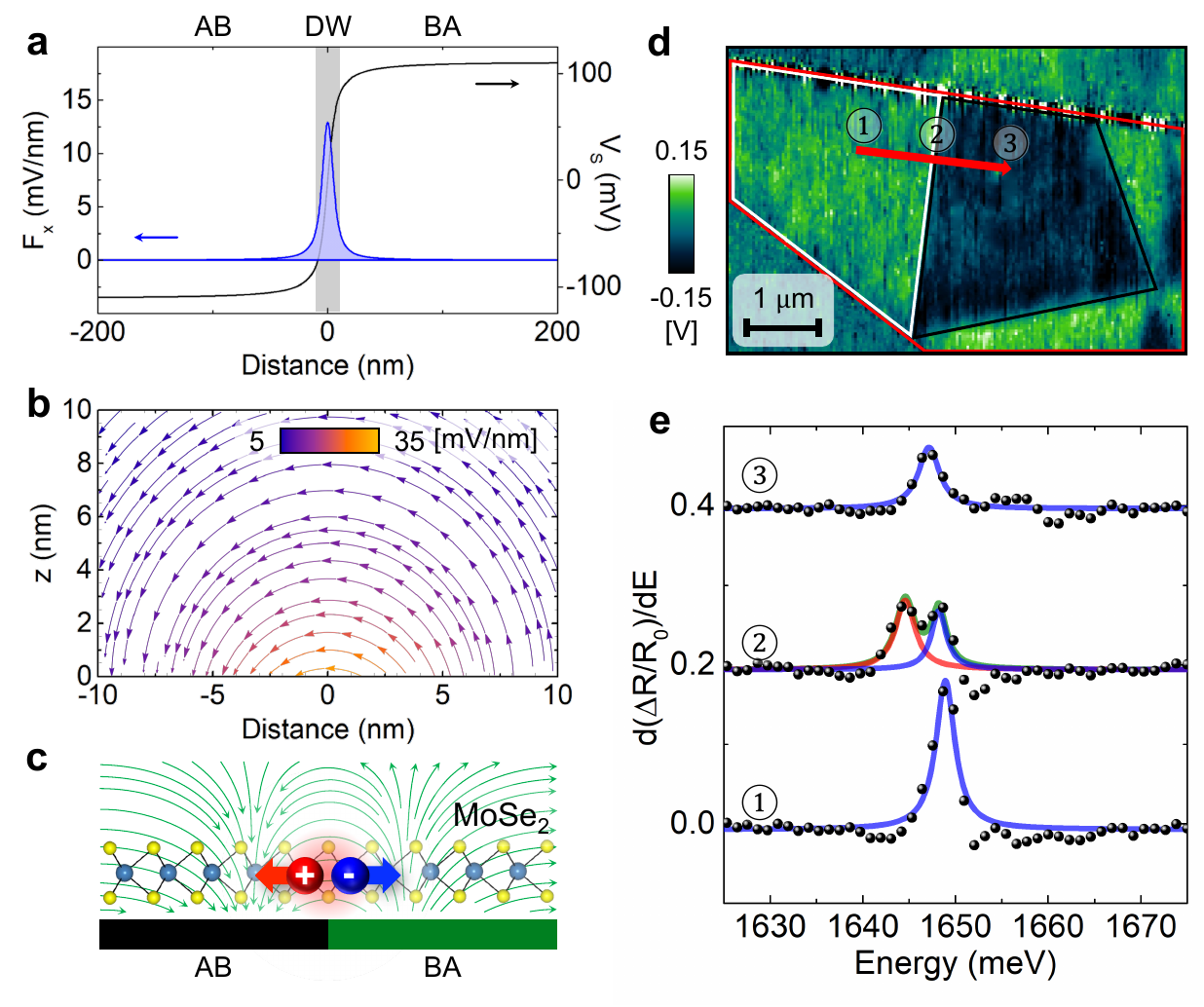}
    \caption{\textbf{Spectral modulation of light emission from MoSe$_2$ due to in-plane E-field at the DWs of a t-hBN substrate.}
    \textbf{a}, Calculated electrostatic potential (black) and in-plane E-field (blue) along a direction perpendicular to a DW.
    \textbf{b}, Side view of E-field lines near the DW. The color represents the field strength.
    \textbf{c}, The in-plane E-field at the DWs separates the electron and hole of an exciton and leads to a Stark-shifted exciton resonance.
    \textbf{d}, KPFM image of a t-hBN substrate forming large domains. Optical spectra are taken from a MoSe$_2$ monolayer at three locations along the red arrow. 
    \textbf{e}, Derivative of reflectance contrast spectra from the MoSe$_2$ monolayer taken at three locations across two opposite and adjacent domains. The numbers correspond to the spots labeled in panel d. One exciton resonance is observed within the AB and BA domains while two resonances are observed at the DW.}
    \label{fig:fig2}
\end{figure}

We examine the in-plane E-field developed at the DWs due to an electrostatic potential drop between adjacent domains (Fig.~\ref{fig:fig2}a). The in-plane E-field was calculated using a supercell size of 500 nm and a top thickness of t-hBN of 4.8 nm. The side view of E-field vectors at the DW is plotted in Fig.~\ref{fig:fig2}b. We hypothesize that the in-plane E-field at the DWs separates the electron and hole without dissociating the excitons in MoSe$_2$ as illustrated in Fig.~\ref{fig:fig2}c. To verify this hypothesis, we conduct experiments by placing a MoSe$_2$ monolayer on a t-hBN substrate with large domains identified via the KPFM image shown in Fig.~\ref{fig:fig2}d. In this particular structure, the separation between the MoSe$_2$ monolayer and the twisted hBN interface, i.e. the top t-hBN thickness, is $\sim$ 12 nm.

Optical reflectivity measurements are taken from three locations across the DW as indicated by the red arrow in Fig.~\ref{fig:fig2}d. The derivatives of reflectance spectra are displayed in Fig.~\ref{fig:fig2}e. The three stacked spectra (bottom to top) are labeled by their respective locations shown in Fig.~\ref{fig:fig2}d. While spectra 1 and 3 taken within a domain feature one exciton resonance, an additional exciton resonance is observed in spectrum 2 taken from a region that overlaps with the DW. We fit these resonances using Lorentzian functions (solid lines in Fig.~\ref{fig:fig2}e). The exciton resonance measured within the AB and BA domains (spectra 1 and 3) exhibits a small energy shift of $\sim$ 1 meV while the higher energy exciton in spectrum 2 lies in-between. Considering the spatial resolution of $\sim$ 1.5 $\mu$m, this intermediate energy likely derives from averaging over excitons residing in the two adjacent and opposite FE domains. We attribute the lower energy exciton in spectrum 2 to exciton Stark shift due to the in-plane E-field at the DW as illustrated in Fig.~\ref{fig:fig2}c. The exciton Stark shift, or the separation between the two resonances is $\sim$ 3 meV in this case.

To establish Stark shift as the mechanism giving rise to the additional resonance at the DWs, we quantitatively evaluate the in-plane E-field, $F_x = \Delta V_\mathrm{S}/d_\mathrm{DW}$ by measuring the potential drop ($\Delta V_\mathrm{S}$) across the DW ($d_\mathrm{DW}$) in multiple samples over regions with different domain sizes (Extended Data Fig.~\ref{fig:fig2}). To the first order, a quadratic relation between the E-field, $F_x$, and exciton Stark shift has been predicted $E_{\mathbf{shift}} = \lvert -\frac{1}{2} \alpha {F_x}^2 \lvert$, where $\alpha$ is the exciton polarizability. We plot the exciton energy shift as a function of the in-plane E-field in Extended Data Fig.~\ref{fig:figs1}. 
Our experiment reports a slightly larger Stark shift or higher exciton polarizability than previous studies~\cite{pedersen_exciton_2016, xiao_exciton-exciton_2023}. This difference may arise from either an underestimated E-field due to a finite tip-sample distance or our specific fitting procedure used to extract the E-field from the KPFM measurements.

\begin{figure}[H]
    \centering
    \includegraphics[width=12cm]{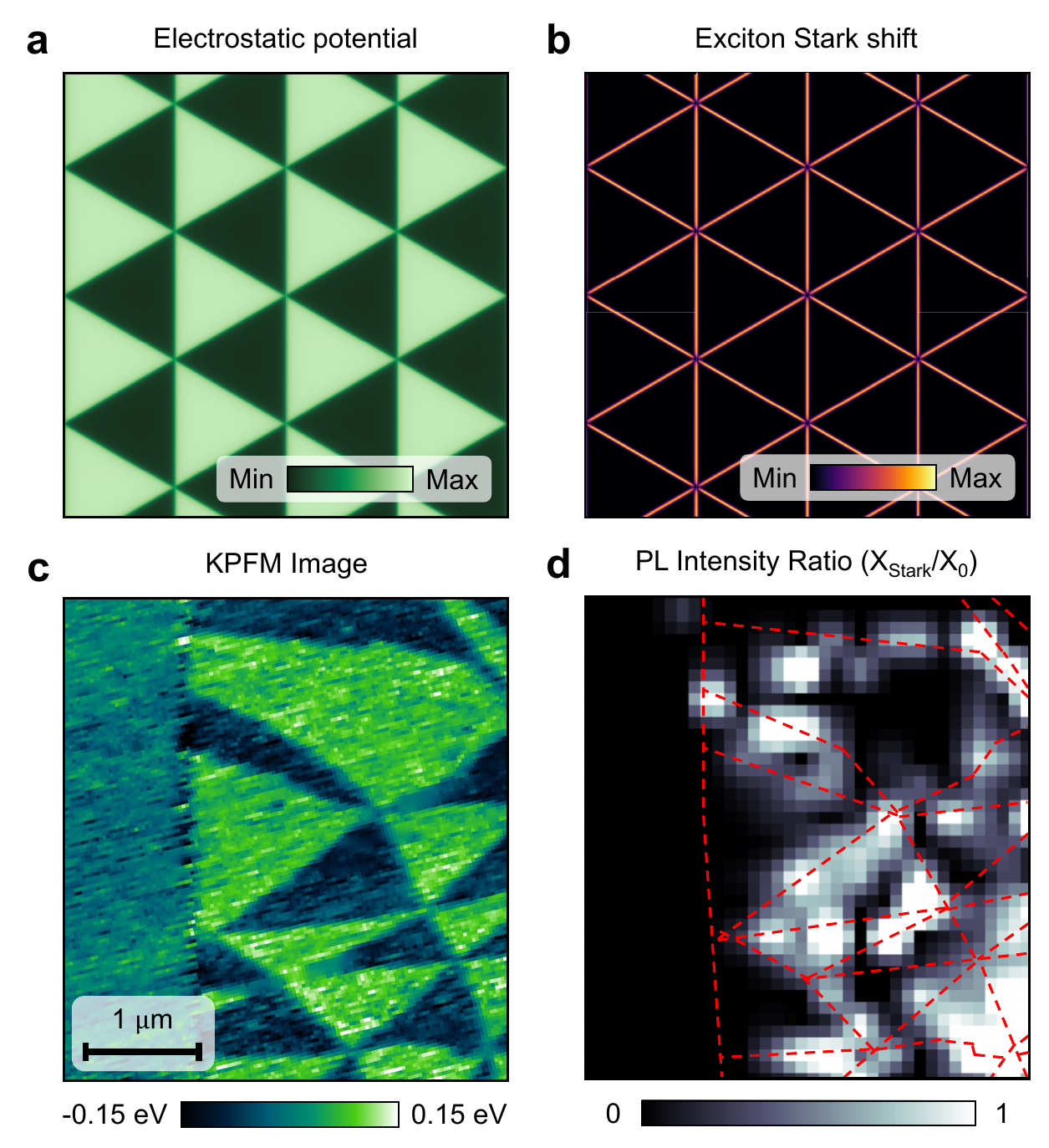}
    \caption{\textbf{Spatial modulation of light emission from MoSe$_2$ by a t-hBN substrate.}
    \textbf{a}, Simulated periodic moir\'e potential on top of a t-hBN substrate.
    \textbf{b}, Simulated exciton Stark shift from a MoSe$_2$ monolayer placed on a t-hBN substrate.
    \textbf{c}, KPFM image of the t-hBN substrate where PL spectra from an adjacent MoSe$_2$ monolayer are collected.
    \textbf{d}, Spatial map of integrated PL intensity ratio of two exciton resonances X$_\textrm{Stark}$/X$_0$. The DWs are marked by red dashed lines.}
    \label{fig:fig3}
\end{figure}

Since the in-plane E-field and the X$_\textrm{Stark}$ are only present at the DWs, one expects not only a spectral but also a spatial modulation of the light emission from the functional semiconductor layer placed on top of the t-hBN with an array of FE domains. We model the electrostatic potential and the corresponding exciton Stark shift induced by an in-plane E-field in the MoSe$_2$/t-hBN heterostructure shown in Fig.~\ref{fig:fig3}a-b (details in Method). We then compare the KPFM image (Fig.~\ref{fig:fig3}c) and confocal photoluminescence image (Fig.~\ref{fig:fig3}d) taken from a MoSe$_2$ monolayer placed on the t-hBN substrate. In Fig.~\ref{fig:fig3}d, we plot the integrated intensity ratio of X$_\textrm{Stark}$ and X$_0$ and observe a spatial correlation between the X$_\mathrm{Stark}$ intensity and the DWs (dashed red lines). Because our measurements are constrained by the optical diffraction limit, we can only image $\sim$ 500 nm or larger domains in our current experimental setup. In principle, light emission modulated by a vdW FE substrate can be used to generate structured light emission on a few nanometer length scales defined by the moir\'e supercells, far below the optical diffraction limit.

\begin{figure}[H]
    \centering
    \includegraphics[width=15cm]{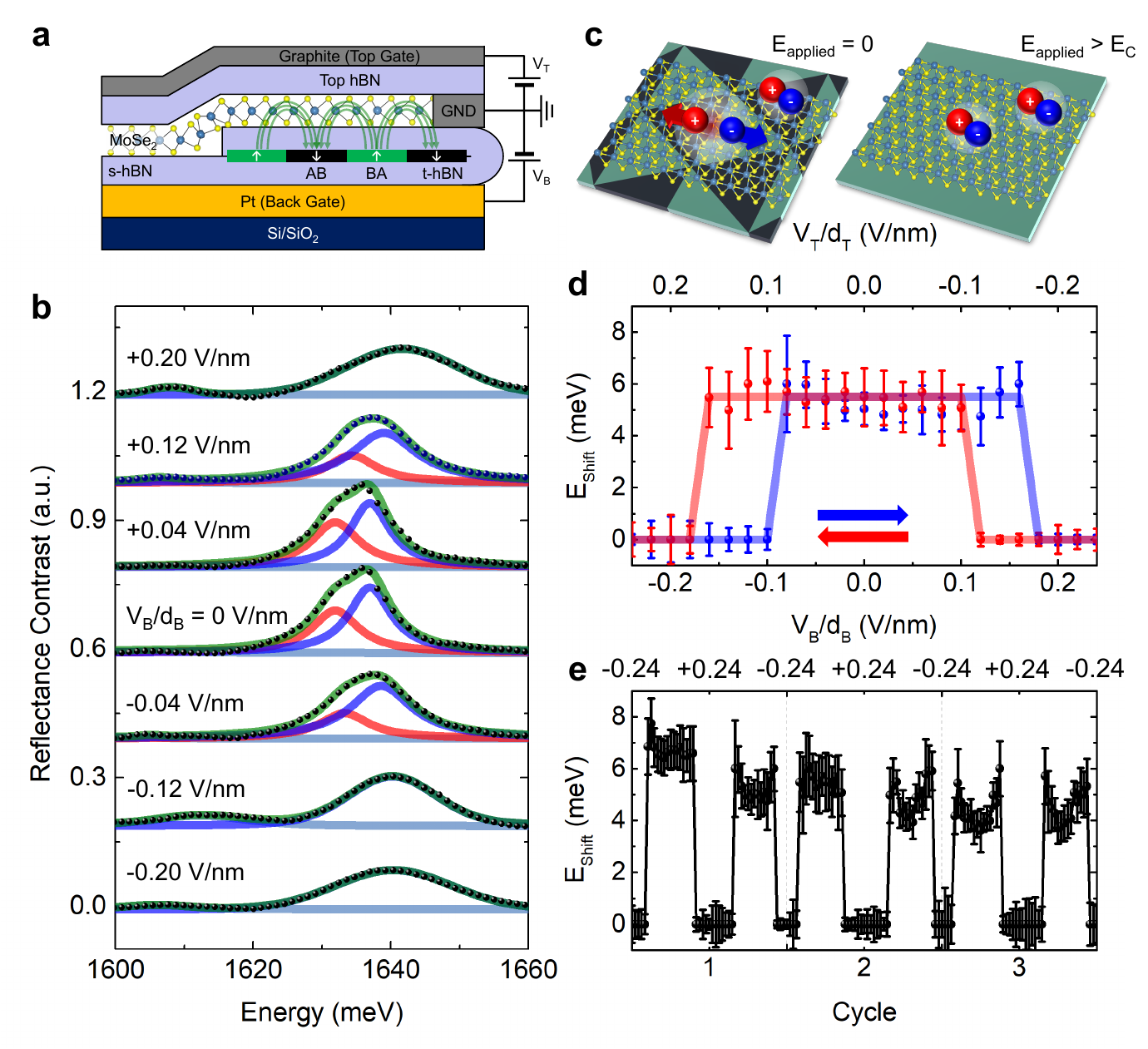}
    \caption{\textbf{Hysteresis of light emission demonstrating the combined functionality of the FE t-hBN substrate with the MoSe$_2$ monolayer.} 
    \textbf{a}, Dual gate device structure. A MoSe$_2$ monolayer is placed on a t-hBN substrate and encapsulated by a top hBN layer. A vertical E-field is applied by adjusting the back-gate voltage V$_B$ and top gate simultaneously while keeping the monolayer undoped.
    \textbf{b}, Reflectance spectra from the MoSe$_2$ monolayer placed on t-hBN substrate at different applied vertical E-fields.
    The exciton Stark shift from the FE domains is observed at V$_B$/d$_B$ = 0 Vnm$^{-1}$. Above V$_B$/d$_B$ = $\pm$ 0.2 Vnm$^{-1}$ (top), the X$_\textrm{Stark}$ peak disappears. Circles are measured data and solid lines are Lorentzian fits.
    \textbf{c}, Illustration of field-driven switching of the FE domains of the t-hBN substrate as manifested via exciton resonances in the adjacent MoSe$_2$ monolayer.
    \textbf{d}, Exciton Stark-shift as a function of the applied E-field. The shift is constant until the resonance abruptly disappears above $\pm$ 0.2 Vnm$^{-1}$. Blue (red) circles are measured exciton Stark-shift as the vertically applied E-field sweeps forward (backward). Hysteresis behavior is characteristic of ferroelectricity from the t-hBN substrate. The solid lines are a guide for the eye. Error bars represent the root mean square errors from the Lorentzian fitting to optical spectra.
    \textbf{e}, Multi-cycle switching of the FE domains manifested by exciton Stark shift. The switching field is $\pm$ 0.24 V/nm.}
    \label{fig:fig4}
\end{figure}

Finally, we demonstrate how an electric gate can be used to modulate light emission from MoSe$_2$ by repeatedly erasing and restoring the FE domains of t-hBN substrates. To accomplish this, we fabricate a dual-gate device structure as illustrated in Fig.~\ref{fig:fig4}a. The device is built on the same stacked layers measured in Fig.~\ref{fig:fig1}c in which the MoSe$_2$ monolayer is separated from the t-hBN interface by 12 nm. We measure a series of reflection spectra (Fig.~\ref{fig:fig4}b) as the vertically applied E-field, $V_\mathrm{B}$/d$_\mathrm{B}$ is varied while keeping the MoSe$_2$ monolayer undoped (i.e. under the condition $V_\mathrm{T}$/d$_\mathrm{T}$ + $V_\mathrm{B}$/d$_\mathrm{B}$ = 0.) In the absence of a vertically applied E-field or a small E-field, exciton doublets are observed due to the exciton Stark shift at the DWs. When the vertically applied E-field is sufficiently large, (i.e., above $V_\mathrm{B}$/d$_\mathrm{B}$ = $\pm$0.2 V/nm), the exciton resonance subject to the Stark shift disappears. We summarize the exciton Stark shift as a function of the vertically applied E-field in Fig.~\ref{fig:fig4}d (see Extended Data Fig.~\ref{fig:figs3} for more details). At a large applied E-field exceeding the critical value, the domain polarization is flipped~\cite{vizner_stern_interfacial_2021, wang_interfacial_2022, ko_operando_2023} such that electrostatic potential from the hBN substrate becomes constant as illustrated in Fig.~\ref{fig:fig4}c. Consequently, the in-plane E-field vanishes and the X$_\textrm{Stark}$ resonance disappears. 

Remarkably, we observe a hysteresis behavior of exciton Stark shift in Fig.~\ref{fig:fig4}d as the gate voltage is swept forward (blue circles) and backward (red circles). This behavior is attributed to the ferroelectricity of the t-hBN substrate. We further demonstrated that this FE switching is reversible and robust over many switching cycles as shown in Fig.~\ref{fig:fig4}e. The bottom gate and top gate are both changed in the range of -6 to 6~V to keep the MoSe$_2$ layer charge neutral. The necessary switching voltage depends on the hBN thickness. The FE domain switching is measured via the exciton Stark shift. Over several cycles, the Stark shifted exciton energy drifts, which likely arise from the DW distortions (Extended Data Fig.~\ref{fig:figs3})~\cite{ko_operando_2023, molino_ferroelectric_2023}. Optical spectroscopy data are typically averaged over several domains except in special cases where we intentionally choose to perform experiments along the DW of a large domain greater than several microns (e.g. Fig.~\ref{fig:fig2}b-c).

We discuss our findings in the context of previous studies. Heterostructures consisting of TMD thin layers placed on conventional FE substrates have been investigated previously~\cite{wen_ferroelectric-driven_2019, wu_programmable_2020, soubelet_charged_2021}. Conventional FE substrates usually have atomically sharp DWs~\cite{gonnissen_direct_2016, xiao_domain_2013} and the in-plane E-field generated by a conventional FE substrate is so strong ($\sim$ 400 mV/nm) that excitons are often dissociated~\cite{soubelet_charged_2021}. In contrast, the in-plane E-field at the surface of t-hBN substrates studied is on the order of $\sim$ 20 mV/nm, which is significantly smaller than the binding energy ($\sim$ 200 meV) of excitons in TMD monolayers~\cite{goryca_revealing_2019}. Furthermore, patterning FE domains requires advanced scanning probes or lithographic tools. The ability to easily create a regular array of FE domains with deep sub-wavelength length scales ( $\sim$ 10 nm) and few-nanometer DW widths are unique advantages of t-hBN in modulating atomically thin semiconductors, establishing a novel platform for engineering active metasurfaces.

In conclusion, we demonstrate that a t-hBN substrate can modulate the light emission from an adjacent semiconductor layer both spectrally and spatially. The in-plane E-field introduced at the FE DWs is sufficiently large to induce a distinct Stark-shifted exciton resonance, similar to that found in semiconductor \textit{p-n} junctions typically created with advanced lithograph tools~\cite{massicotte_dissociation_2018, thureja_electrically_2022}. The light emission can be further modulated by an electric gate that erases and restores the FE domains in the t-hBN. The hysteresis behavior observed in optical spectra demonstrates that the FE property of the t-hBN substrate is successfully combined with the light emission functionality of the semiconductor layer. Our findings open the exciting possibilities for designing new metasurfaces and optoelectronic devices based on FE-hBN substrates. For example, FE tuning of exciton resonances may facilitate coupling to optical cavities, enabling polariton-based photonic devices.

\section*{Methods}

\subsection{Sample preparation:} 
We exfoliated hBN flakes using scotch tape onto 285~nm SiO$_2$/Si substrates. After choosing a target hBN flake via microscope imaging, nitrogen gas is blown to facilitate the folding process. Samples typically fold along either zigzag or armchair directions. After stacking or folding, the samples are annealed up to 400$\degree$C for 4 hours under vacuum $\sim$ 10$^{-7}$ Torr to increase interface bonding by removing polymer residue. For the MoSe$_2$/t-hBN device structure, pre-patterned Pt/Ti is prepared by photolithography. Then t-hBN substrate is transferred using a 15\% PPC solution and dissolved into anisole. Then contact graphite, MoSe$_2$ monolayer, top hBN, and top gate graphite are transferred in the same way.

\subsection{KPFM measurements:}
Kelvin probe force microscopy (KPFM) measurements were performed using SmartSPM (Horiba) in two-pass frequency modulation KPFM (FM-KPFM) mode. All the data were taken by FM-KPFM mode. We used Pt-coated conductive cantilever probes with a nominal resonance frequency of 70~kHz and a spring constant of 2~N/m (OPUS 240AC-PP, Mikromasch), and the Au-coated cantilevers with supersharp diamond-like carbon tips with a nominal resonance frequency of 150~kHz and a spring constant of 5~N/m (BudgetSensors SHR150). In the FM-KPFM mode of SmartSPM, the resonance frequency shift of the mechanically excited oscillations (amplitude 20~nm), which are caused by the electrostatic force gradient with respect to the tip-sample distance, is detected via the phase of the cantilever oscillations. 
The amplitude of the modulation of the phase, which is caused by applying an ac voltage (3~V, 1~kHz), is proportional to the difference between the applied dc bias voltage between the contact potential difference (CPD) and is fed into a feedback controller to nullify the electrostatic force.

\subsection{Optical spectroscopy measurements:}
For optical reflectivity measurements at 14~K, a compact stabilized broadband light source was focused to a spot size of $\sim$ 2 $\mu$m in diameter using a 100 X microscope objective. For photoluminescence measurements, a He-Ne laser was used, and the excitation power was kept below 5 $\mu$W to avoid local heating. The spatial mapping of photoluminescence was conducted in confocal geometry.

\subsection{Computational methods:}

Extending the formula in \cite{zhao_universal_2021} to a non-rigid twist region by considering all the reciprocal vectors, we can calculate the static electric potential at any position,
\begin{equation}
    \begin{aligned}
        & V(\bm{R}, z)\approx\sum_{\bm{G}} \frac{\tilde{P}(\bm{G})}{2 \varepsilon_0} \mathrm{e}^{\mathrm{i} \bm{G} \cdot \bm{R}} \mathrm{e}^{-|\bm{G}||z|} \\
    \end{aligned}\label{eq-potential}
\end{equation}
where \(\tilde{P}(\bm{G})\) is the Fourier components of the electric dipole \(P(\bm{R})=\int \Delta \rho (\bm{R},z)z dz\), with \(\Delta \rho\) denoting the differential charge density, and \(\bm{G}\) is the moir\'e reciprocal vectors.
The strong lattice reconstruction in the marginally twisted hBN results in a narrow domain wall (DW) between AB and BA stacking. Former ab initio study~\cite{lebedev_interlayer_2016} shows the DW width of strained bilayer hBN is around \(8\sim 10\) nm. Thus, we calculated the electric potential and the electric field in Fig. \ref{fig:fig1}, using Eqn. \eqref{eq-potential} with triangular modified \(P(\bm{R})\) and 8 nm DW width.

\section*{References}
\bibliography{Stark-MoSe2-thBN}

\begin{thebibliography}{10}
\expandafter\ifx\csname url\endcsname\relax
  \def\url#1{\texttt{#1}}\fi
\expandafter\ifx\csname urlprefix\endcsname\relax\def\urlprefix{URL }\fi
\providecommand{\bibinfo}[2]{#2}
\providecommand{\eprint}[2][]{\url{#2}}

\bibitem{seyler_electrical_2015}
\bibinfo{author}{Seyler, K.~L.} \emph{et~al.}
\newblock \bibinfo{title}{Electrical control of second-harmonic generation in a
  {WSe2} monolayer transistor}.
\newblock \emph{\bibinfo{journal}{Nature Nanotechnology}}
  \textbf{\bibinfo{volume}{10}}, \bibinfo{pages}{407--411}
  (\bibinfo{year}{2015}).

\bibitem{sun_optical_2016}
\bibinfo{author}{Sun, Z.}, \bibinfo{author}{Martinez, A.} \&
  \bibinfo{author}{Wang, F.}
\newblock \bibinfo{title}{Optical modulators with {2D} layered materials}.
\newblock \emph{\bibinfo{journal}{Nature Photonics}}
  \textbf{\bibinfo{volume}{10}}, \bibinfo{pages}{227--238}
  (\bibinfo{year}{2016}).

\bibitem{lopez-sanchez_ultrasensitive_2013}
\bibinfo{author}{Lopez-Sanchez, O.}, \bibinfo{author}{Lembke, D.},
  \bibinfo{author}{Kayci, M.}, \bibinfo{author}{Radenovic, A.} \&
  \bibinfo{author}{Kis, A.}
\newblock \bibinfo{title}{Ultrasensitive photodetectors based on monolayer
  {MoS2}}.
\newblock \emph{\bibinfo{journal}{Nature Nanotechnology}}
  \textbf{\bibinfo{volume}{8}}, \bibinfo{pages}{497--501}
  (\bibinfo{year}{2013}).

\bibitem{baugher_optoelectronic_2014}
\bibinfo{author}{Baugher, B. W.~H.}, \bibinfo{author}{Churchill, H. O.~H.},
  \bibinfo{author}{Yang, Y.} \& \bibinfo{author}{Jarillo-Herrero, P.}
\newblock \bibinfo{title}{Optoelectronic devices based on electrically tunable
  p–n diodes in a monolayer dichalcogenide}.
\newblock \emph{\bibinfo{journal}{Nature Nanotechnology}}
  \textbf{\bibinfo{volume}{9}}, \bibinfo{pages}{262--267}
  (\bibinfo{year}{2014}).

\bibitem{ross_electrically_2014}
\bibinfo{author}{Ross, J.~S.} \emph{et~al.}
\newblock \bibinfo{title}{Electrically tunable excitonic light-emitting diodes
  based on monolayer {WSe2} p–n junctions}.
\newblock \emph{\bibinfo{journal}{Nature Nanotechnology}}
  \textbf{\bibinfo{volume}{9}}, \bibinfo{pages}{268--272}
  (\bibinfo{year}{2014}).

\bibitem{withers_light-emitting_2015}
\bibinfo{author}{Withers, F.} \emph{et~al.}
\newblock \bibinfo{title}{Light-emitting diodes by band-structure engineering
  in van der {Waals} heterostructures}.
\newblock \emph{\bibinfo{journal}{Nature Materials}}
  \textbf{\bibinfo{volume}{14}}, \bibinfo{pages}{301--306}
  (\bibinfo{year}{2015}).

\bibitem{wilson_excitons_2021}
\bibinfo{author}{Wilson, N.~P.}, \bibinfo{author}{Yao, W.},
  \bibinfo{author}{Shan, J.} \& \bibinfo{author}{Xu, X.}
\newblock \bibinfo{title}{Excitons and emergent quantum phenomena in stacked
  {2D} semiconductors}.
\newblock \emph{\bibinfo{journal}{Nature}} \textbf{\bibinfo{volume}{599}},
  \bibinfo{pages}{383--392} (\bibinfo{year}{2021}).

\bibitem{huang_excitons_2022}
\bibinfo{author}{Huang, D.}, \bibinfo{author}{Choi, J.}, \bibinfo{author}{Shih,
  C.-K.} \& \bibinfo{author}{Li, X.}
\newblock \bibinfo{title}{Excitons in semiconductor moiré superlattices}.
\newblock \emph{\bibinfo{journal}{Nature Nanotechnology}}
  \textbf{\bibinfo{volume}{17}}, \bibinfo{pages}{227--238}
  (\bibinfo{year}{2022}).

\bibitem{mak_semiconductor_2022}
\bibinfo{author}{Mak, K.~F.} \& \bibinfo{author}{Shan, J.}
\newblock \bibinfo{title}{Semiconductor moiré materials}.
\newblock \emph{\bibinfo{journal}{Nature Nanotechnology}}
  \textbf{\bibinfo{volume}{17}}, \bibinfo{pages}{686--695}
  (\bibinfo{year}{2022}).

\bibitem{choi_moire_2020}
\bibinfo{author}{Choi, J.} \emph{et~al.}
\newblock \bibinfo{title}{Moiré potential impedes interlayer exciton diffusion
  in van der {Waals} heterostructures}.
\newblock \emph{\bibinfo{journal}{Science Advances}}
  \textbf{\bibinfo{volume}{6}}, \bibinfo{pages}{eaba8866}
  (\bibinfo{year}{2020}).

\bibitem{choi_twist_2021}
\bibinfo{author}{Choi, J.} \emph{et~al.}
\newblock \bibinfo{title}{Twist {Angle}-{Dependent} {Interlayer} {Exciton}
  {Lifetimes} in van der {Waals} {Heterostructures}}.
\newblock \emph{\bibinfo{journal}{Physical Review Letters}}
  \textbf{\bibinfo{volume}{126}}, \bibinfo{pages}{047401}
  (\bibinfo{year}{2021}).

\bibitem{tran_quantum_2016}
\bibinfo{author}{Tran, T.~T.}, \bibinfo{author}{Bray, K.},
  \bibinfo{author}{Ford, M.~J.}, \bibinfo{author}{Toth, M.} \&
  \bibinfo{author}{Aharonovich, I.}
\newblock \bibinfo{title}{Quantum emission from hexagonal boron nitride
  monolayers}.
\newblock \emph{\bibinfo{journal}{Nature Nanotechnology}}
  \textbf{\bibinfo{volume}{11}}, \bibinfo{pages}{37--41}
  (\bibinfo{year}{2016}).

\bibitem{grosso_tunable_2017}
\bibinfo{author}{Grosso, G.} \emph{et~al.}
\newblock \bibinfo{title}{Tunable and high-purity room temperature
  single-photon emission from atomic defects in hexagonal boron nitride}.
\newblock \emph{\bibinfo{journal}{Nature Communications}}
  \textbf{\bibinfo{volume}{8}}, \bibinfo{pages}{705} (\bibinfo{year}{2017}).

\bibitem{maity_hexagonal_2018}
\bibinfo{author}{Maity, A.}, \bibinfo{author}{Grenadier, S.~J.},
  \bibinfo{author}{Li, J.}, \bibinfo{author}{Lin, J.~Y.} \&
  \bibinfo{author}{Jiang, H.~X.}
\newblock \bibinfo{title}{Hexagonal boron nitride neutron detectors with high
  detection efficiencies}.
\newblock \emph{\bibinfo{journal}{Journal of Applied Physics}}
  \textbf{\bibinfo{volume}{123}}, \bibinfo{pages}{044501}
  (\bibinfo{year}{2018}).

\bibitem{dai_tunable_2014}
\bibinfo{author}{Dai, S.} \emph{et~al.}
\newblock \bibinfo{title}{Tunable {Phonon} {Polaritons} in {Atomically} {Thin}
  van der {Waals} {Crystals} of {Boron} {Nitride}}.
\newblock \emph{\bibinfo{journal}{Science}} \textbf{\bibinfo{volume}{343}},
  \bibinfo{pages}{1125--1129} (\bibinfo{year}{2014}).

\bibitem{caldwell_sub-diffractional_2014}
\bibinfo{author}{Caldwell, J.~D.} \emph{et~al.}
\newblock \bibinfo{title}{Sub-diffractional volume-confined polaritons in the
  natural hyperbolic material hexagonal boron nitride}.
\newblock \emph{\bibinfo{journal}{Nature Communications}}
  \textbf{\bibinfo{volume}{5}}, \bibinfo{pages}{5221} (\bibinfo{year}{2014}).

\bibitem{yasuda_stacking-engineered_2021}
\bibinfo{author}{Yasuda, K.}, \bibinfo{author}{Wang, X.},
  \bibinfo{author}{Watanabe, K.}, \bibinfo{author}{Taniguchi, T.} \&
  \bibinfo{author}{Jarillo-Herrero, P.}
\newblock \bibinfo{title}{Stacking-engineered ferroelectricity in bilayer boron
  nitride}.
\newblock \emph{\bibinfo{journal}{Science}} \textbf{\bibinfo{volume}{372}},
  \bibinfo{pages}{1458--1462} (\bibinfo{year}{2021}).

\bibitem{vizner_stern_interfacial_2021}
\bibinfo{author}{Vizner~Stern, M.} \emph{et~al.}
\newblock \bibinfo{title}{Interfacial ferroelectricity by van der {Waals}
  sliding}.
\newblock \emph{\bibinfo{journal}{Science}} \textbf{\bibinfo{volume}{372}},
  \bibinfo{pages}{1462--1466} (\bibinfo{year}{2021}).

\bibitem{zhao_universal_2021}
\bibinfo{author}{Zhao, P.}, \bibinfo{author}{Xiao, C.} \& \bibinfo{author}{Yao,
  W.}
\newblock \bibinfo{title}{Universal superlattice potential for {2D} materials
  from twisted interface inside h-{BN} substrate}.
\newblock \emph{\bibinfo{journal}{npj 2D Materials and Applications}}
  \textbf{\bibinfo{volume}{5}}, \bibinfo{pages}{1--7} (\bibinfo{year}{2021}).

\bibitem{wu_programmable_2020}
\bibinfo{author}{Wu, G.} \emph{et~al.}
\newblock \bibinfo{title}{Programmable transition metal dichalcogenide
  homojunctions controlled by nonvolatile ferroelectric domains}.
\newblock \emph{\bibinfo{journal}{Nature Electronics}}
  \textbf{\bibinfo{volume}{3}}, \bibinfo{pages}{43--50} (\bibinfo{year}{2020}).

\bibitem{forbes_structured_2021}
\bibinfo{author}{Forbes, A.}, \bibinfo{author}{de~Oliveira, M.} \&
  \bibinfo{author}{Dennis, M.~R.}
\newblock \bibinfo{title}{Structured light}.
\newblock \emph{\bibinfo{journal}{Nature Photonics}}
  \textbf{\bibinfo{volume}{15}}, \bibinfo{pages}{253--262}
  (\bibinfo{year}{2021}).

\bibitem{kim_electrostatic_2024}
\bibinfo{author}{Kim, D.~S.} \emph{et~al.}
\newblock \bibinfo{title}{Electrostatic moiré potential from twisted hexagonal
  boron nitride layers}.
\newblock \emph{\bibinfo{journal}{Nature Materials}}
  \textbf{\bibinfo{volume}{23}}, \bibinfo{pages}{65--70}
  (\bibinfo{year}{2024}).

\bibitem{pedersen_exciton_2016}
\bibinfo{author}{Pedersen, T.~G.}
\newblock \bibinfo{title}{Exciton {Stark} shift and electroabsorption in
  monolayer transition-metal dichalcogenides}.
\newblock \emph{\bibinfo{journal}{Physical Review B}}
  \textbf{\bibinfo{volume}{94}}, \bibinfo{pages}{125424}
  (\bibinfo{year}{2016}).

\bibitem{xiao_exciton-exciton_2023}
\bibinfo{author}{Xiao, K.} \emph{et~al.}
\newblock \bibinfo{title}{Exciton-exciton {Interaction} in {Monolayer}
  {MoSe}\$\_2\$ from {Mutual} {Screening} of {Coulomb} {Binding}}
  (\bibinfo{year}{2023}).
\newblock \urlprefix\url{http://arxiv.org/abs/2308.14362}.

\bibitem{wang_interfacial_2022}
\bibinfo{author}{Wang, X.} \emph{et~al.}
\newblock \bibinfo{title}{Interfacial ferroelectricity in rhombohedral-stacked
  bilayer transition metal dichalcogenides}.
\newblock \emph{\bibinfo{journal}{Nature Nanotechnology}}
  \textbf{\bibinfo{volume}{17}}, \bibinfo{pages}{367--371}
  (\bibinfo{year}{2022}).

\bibitem{ko_operando_2023}
\bibinfo{author}{Ko, K.} \emph{et~al.}
\newblock \bibinfo{title}{Operando electron microscopy investigation of polar
  domain dynamics in twisted van der {Waals} homobilayers}.
\newblock \emph{\bibinfo{journal}{Nature Materials}}
  \textbf{\bibinfo{volume}{22}}, \bibinfo{pages}{992--998}
  (\bibinfo{year}{2023}).

\bibitem{molino_ferroelectric_2023}
\bibinfo{author}{Molino, L.} \emph{et~al.}
\newblock \bibinfo{title}{Ferroelectric {Switching} at {Symmetry}-{Broken}
  {Interfaces} by {Local} {Control} of {Dislocations} {Networks}}.
\newblock \emph{\bibinfo{journal}{Advanced Materials}}
  \textbf{\bibinfo{volume}{35}}, \bibinfo{pages}{2207816}
  (\bibinfo{year}{2023}).

\bibitem{wen_ferroelectric-driven_2019}
\bibinfo{author}{Wen, B.} \emph{et~al.}
\newblock \bibinfo{title}{Ferroelectric-{Driven} {Exciton} and {Trion}
  {Modulation} in {Monolayer} {Molybdenum} and {Tungsten} {Diselenides}}.
\newblock \emph{\bibinfo{journal}{ACS Nano}} \textbf{\bibinfo{volume}{13}},
  \bibinfo{pages}{5335--5343} (\bibinfo{year}{2019}).

\bibitem{soubelet_charged_2021}
\bibinfo{author}{Soubelet, P.} \emph{et~al.}
\newblock \bibinfo{title}{Charged {Exciton} {Kinetics} in {Monolayer} {MoSe2}
  near {Ferroelectric} {Domain} {Walls} in {Periodically} {Poled} {LiNbO3}}.
\newblock \emph{\bibinfo{journal}{Nano Letters}} \textbf{\bibinfo{volume}{21}},
  \bibinfo{pages}{959--966} (\bibinfo{year}{2021}).

\bibitem{gonnissen_direct_2016}
\bibinfo{author}{Gonnissen, J.} \emph{et~al.}
\newblock \bibinfo{title}{Direct {Observation} of {Ferroelectric} {Domain}
  {Walls} in {LiNbO3}: {Wall}-{Meanders}, {Kinks}, and {Local} {Electric}
  {Charges}}.
\newblock \emph{\bibinfo{journal}{Advanced Functional Materials}}
  \textbf{\bibinfo{volume}{26}}, \bibinfo{pages}{7599--7604}
  (\bibinfo{year}{2016}).

\bibitem{xiao_domain_2013}
\bibinfo{author}{Xiao, Z.}, \bibinfo{author}{Poddar, S.},
  \bibinfo{author}{Ducharme, S.} \& \bibinfo{author}{Hong, X.}
\newblock \bibinfo{title}{Domain wall roughness and creep in nanoscale
  crystalline ferroelectric polymers}.
\newblock \emph{\bibinfo{journal}{Applied Physics Letters}}
  \textbf{\bibinfo{volume}{103}}, \bibinfo{pages}{112903}
  (\bibinfo{year}{2013}).

\bibitem{goryca_revealing_2019}
\bibinfo{author}{Goryca, M.} \emph{et~al.}
\newblock \bibinfo{title}{Revealing exciton masses and dielectric properties of
  monolayer semiconductors with high magnetic fields}.
\newblock \emph{\bibinfo{journal}{Nature Communications}}
  \textbf{\bibinfo{volume}{10}}, \bibinfo{pages}{4172} (\bibinfo{year}{2019}).

\bibitem{massicotte_dissociation_2018}
\bibinfo{author}{Massicotte, M.} \emph{et~al.}
\newblock \bibinfo{title}{Dissociation of two-dimensional excitons in monolayer
  {WSe2}}.
\newblock \emph{\bibinfo{journal}{Nature Communications}}
  \textbf{\bibinfo{volume}{9}}, \bibinfo{pages}{1633} (\bibinfo{year}{2018}).

\bibitem{thureja_electrically_2022}
\bibinfo{author}{Thureja, D.} \emph{et~al.}
\newblock \bibinfo{title}{Electrically tunable quantum confinement of neutral
  excitons}.
\newblock \emph{\bibinfo{journal}{Nature}} \textbf{\bibinfo{volume}{606}},
  \bibinfo{pages}{298--304} (\bibinfo{year}{2022}).

\bibitem{lebedev_interlayer_2016}
\bibinfo{author}{Lebedev, A.~V.}, \bibinfo{author}{Lebedeva, I.~V.},
  \bibinfo{author}{Knizhnik, A.~A.} \& \bibinfo{author}{Popov, A.~M.}
\newblock \bibinfo{title}{Interlayer interaction and related properties of
  bilayer hexagonal boron nitride: ab initio study}.
\newblock \emph{\bibinfo{journal}{RSC Advances}} \textbf{\bibinfo{volume}{6}},
  \bibinfo{pages}{6423--6435} (\bibinfo{year}{2016}).

\end{thebibliography}

\section*{Data availability}
The data that support the plots within this paper and other findings of this study are available from the corresponding authors upon reasonable request. Source data are provided with this paper.

\section*{Acknowledgements} The experiments are primarily supported by NSF MRSEC DMR-2308817 (D. S. K.), Army Research Office W911NF-23-1-0364 (K.L.), and NSF ECCS-2130552 (Z. L. and M.A.). D. S. K. is partially supported by NSF Designing Materials to Revolutionize and Engineer our Future (DMREF) program via grants DMR-2118806. X.L. gratefully acknowledges support from the Welch Foundation Chair F-0014.
C.X. and W.Y. acknowledge support by RGC of HKSAR (AoE/P-701/20, HKU SRFS2122-7S05), and the New Cornerstone Science Foundation.
The work by R. C. D. and R. M-L. is supported by the NSF Partnership for Research and Education in Materials (PREM) (NSF award DMR-2122041).
Y. M. acknowledges the support by the NSF CAREER (DMR-2044920) and NSF MRI (DMR-2117438) grants.
The collaboration between X. L. and C. S. are enabled by the National Science Foundation through the Center for Dynamics and Control of Materials: an NSF MRSEC under Cooperative Agreement No.~DMR-1720595 and DMR-2308817, which also supported the user facility where part of the experiments were performed. C.K.S. acknowledges support from the NSF grant nos. DMR-2219610, the Welch Foundation F-2164, and the US Air Force grant no. FA2386-21-1-4061
K. W. and T. T. acknowledge support from the JSPS KAKENHI (Grant Numbers 20H00354, 21H05233 and 23H02052) and World Premier International Research Center Initiative (WPI), MEXT, Japan.

\section*{Author contributions}
D. S. K. and X. L. conceived the project. D. S. K. fabricated samples with assistance from K. L.. D. S. K. carried out optical measurements with H. A. and Z. L.'s help, and R. C. D. and R. M-L. performed AFM and KPFM measurements. Electrical gating measurements were done by D. S. K. with contributions from H. K. and Z. L.. K. W. and T. T. synthesized hBN single crystals. C. X., and W. Y. proposed the theoretical model and performed the calculation. D. S. K. analyzed the data with contributions from C. X.. The first draft of the manuscript was written by D. S. K., C. K. S., Y. M., W. Y., and X. L.. All authors contributed to discussions.

\section*{Competing interests}
The authors declare no competing interests.

\newpage
\section*{Figures}

\renewcommand{\baselinestretch}{1}

\newpage

\newpage

\newpage

\bibdata{Stark-MoSe2-thBN.bib}


\setcounter{figure}{0}
\setcounter{table}{0}

\newpage
\section*{Extended Data}

\makeatletter
\let\saved@includegraphics\includegraphics
\renewenvironment*{figure}{\@float{figure}}{\end@float}
\renewenvironment*{table}{\@float{table}}{\end@float}

\renewcommand{\figurename}{Extended Data Fig.}
\renewcommand{\tablename}{Extended Data Table.}

\renewcommand{\baselinestretch}{1}

\begin{figure}[H]
    \centering
    \includegraphics[width=\linewidth]{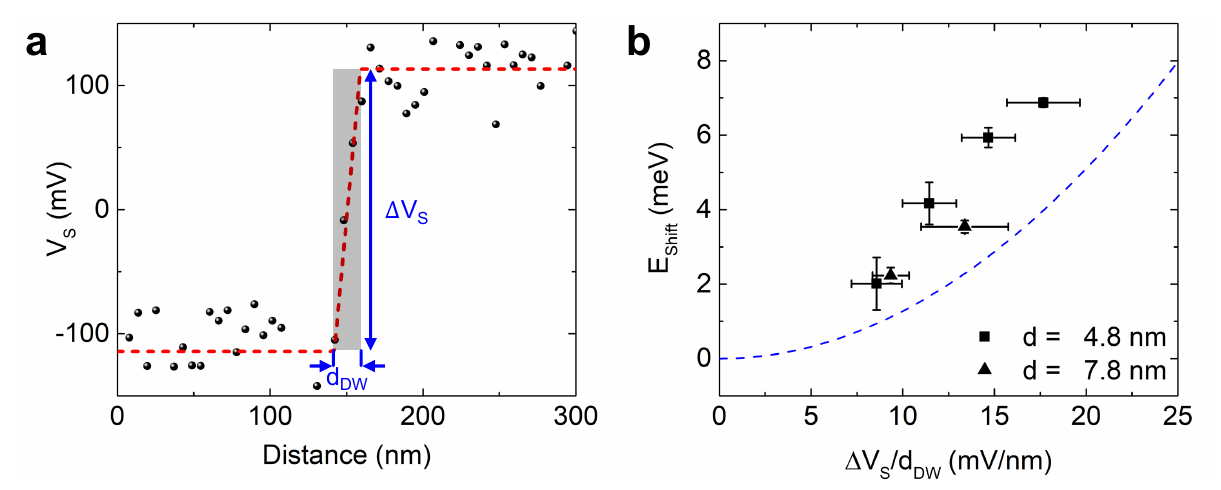}
    \caption{\textbf{Dependence of Stark shift on in-plane E-field at the FE DWs at t-hBN substrate.}
    \textbf{a}, Electrostatic potential drops ($\Delta$V$_\textrm{S}$ indicated by the vertical blue arrow) across the DW (d$_{DW}$) of the t-hBN substrate, producing an in-plane E-field and causing exciton Stark shift. The grey strip indicates a finite DW thickness (d$_{DW}$). The dashed red line is a guide for the eye.
    \textbf{b}, Summary of the exciton Stark shift as a function of the in-plane E-field, extracted from the KPFM measurements via $\Delta$V$_\textrm{S}$/d$_\textrm{DW}$. KPFM data are taken from two bare t-hBN substrates with different hBN thicknesses, $d$. The dashed blue line is the calculated exciton Stark shift. The error bars in the horizontal direction derive from both the $\Delta V_\mathrm{S}$ and $d_\mathrm{DW}$. The error bars in the vertical direction are fitting errors in extracting the exciton energy splitting for spectra taken across DWs.}
    \label{fig:figs1}
\end{figure}

\newpage
\begin{figure}[H]
    \centering
    \includegraphics[width=14cm]{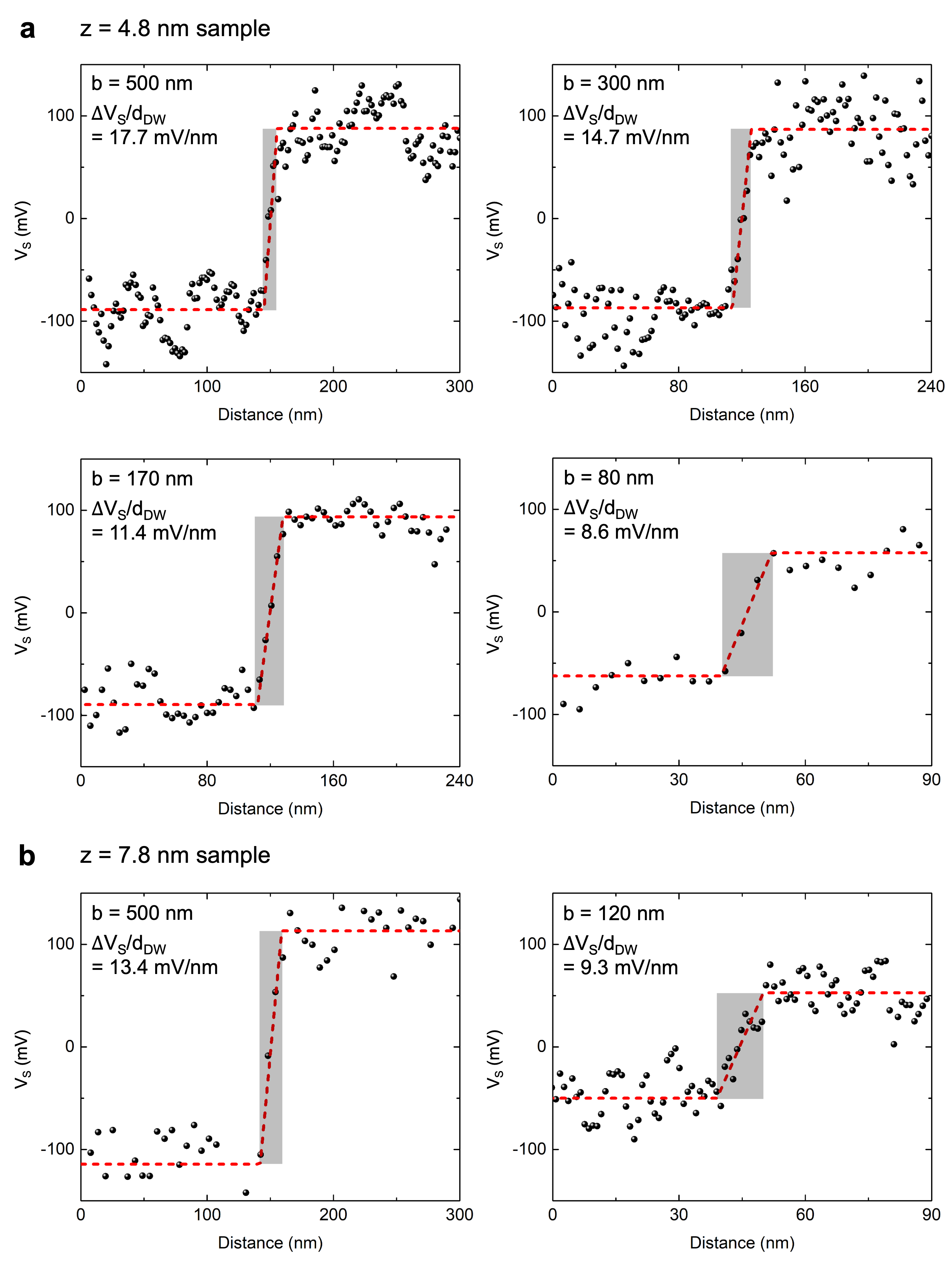}
    \caption{\textbf{Evaluation of the in-plane E-field for additional data points in Extended Data Fig. 1b.}
    \textbf{a, b}, KPFM line cuts showing electric potential drops across the DW of the t-hBN substrate with a thickness of (\textbf{a}) 4.8 nm and (\textbf{b}) 7.8 nm.}
    \label{fig:figs2}
\end{figure}

\newpage
\begin{figure}[H]
    \centering
    \includegraphics[width=\linewidth]{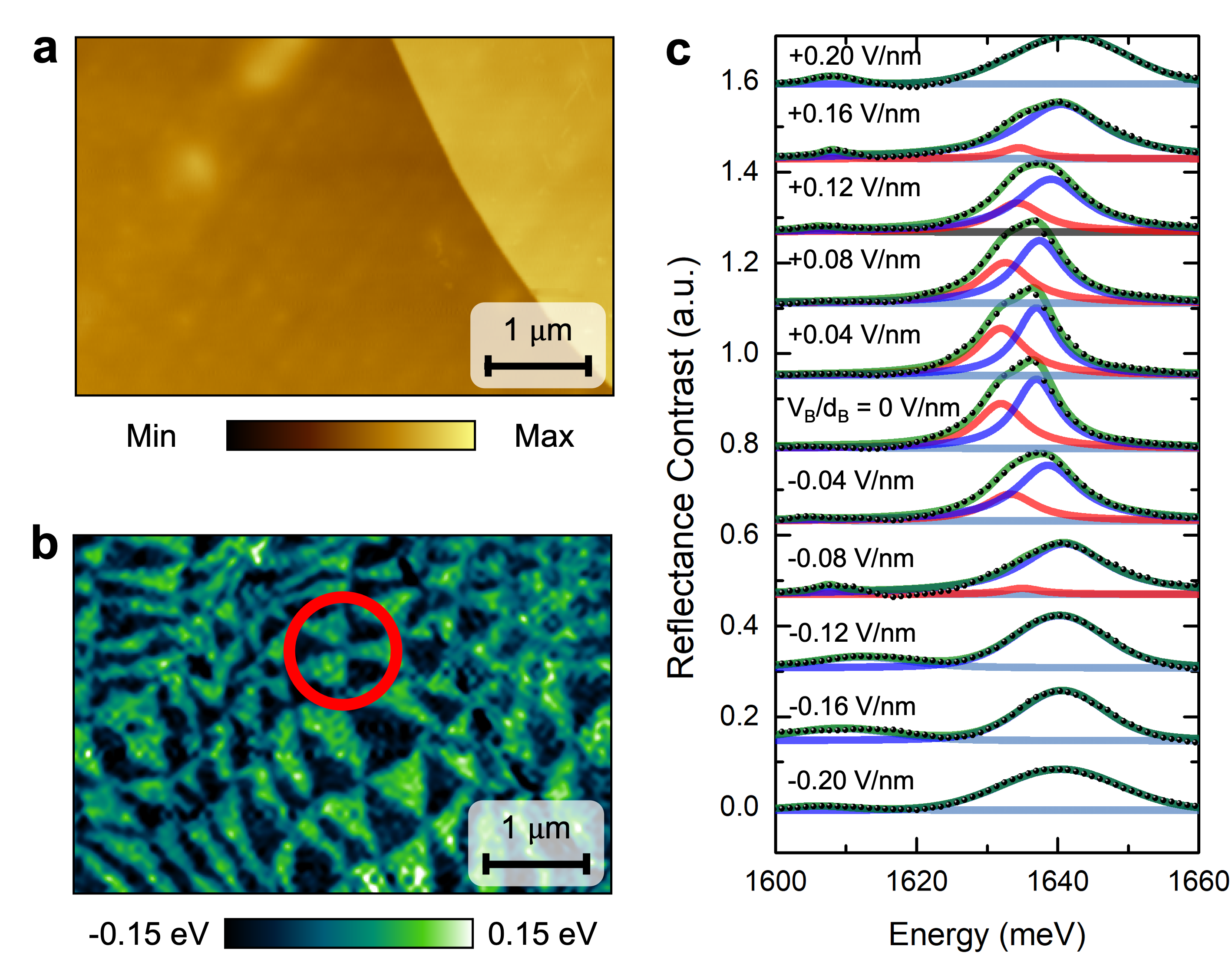}
    \caption{\textbf{Additional reflectance spectra from the MoSe$_2$/t-hBN heterostructure used in Fig. 3d.}
    \textbf{a}, Topography of a t-hBN substrate.
    \textbf{b}, Corresponding KPFM image. The red circle indicates the location where the reflectance spectra were taken.
    \textbf{c}, Reflectance spectra from the MoSe$_2$ monolayer as the applied vertical E-field is varied.
    }
    \label{fig:figs3}
\end{figure}

\bibdata{Stark-MoSe2-thBN.bib}

\end{document}